\documentclass[english,aps,manuscript]{article}
\usepackage[LGR,T1]{fontenc}
\usepackage[latin9]{inputenc}
\usepackage{float}
\usepackage{textcomp}
\usepackage{amstext}
\usepackage{graphicx}
\usepackage{setspace}
\usepackage{wasysym}
\makeatletter

\DeclareRobustCommand{\greektext}{%
  \fontencoding{LGR}\selectfont\def\encodingdefault{LGR}}
\DeclareRobustCommand{\textgreek}[1]{\leavevmode{\greektext #1}}
\ProvideTextCommand{\~}{LGR}[1]{\char126#1}

\newcommand{\lyxaddress}[1]{
\par {\raggedright #1
\vspace{1.4em}
\noindent\par}
}

\makeatother

\usepackage{babel}
\begin{document}

\title{Two Flavour Neutrino Oscillation in Matter and Quantum Entanglement}

\author{Bipin Singh Koranga and Baktiar Wasir Farooq}
\maketitle

\lyxaddress{Department of Physics, Kirori Mal College (University of Delhi,)
Delhi-110007, India.}
\begin{abstract}
In this article, we investigate the entanglement entropy for neutrino
oscillations when neutrino propogate in matter, utilising Von Neumann
entropy. We discuss two flavour neutrino oscillation in vaccum and
matter. We demonstrate statistically that, depending on the length
of oscillation for each energy, the entanglement entropy for the succeeding
periods of the two-flavor neutrino oscillations in matter. 
\end{abstract}

\section{Introduction}

In neutrino physics, quantum entanglement in neutrino oscillations
have garnered significant attention recently {[}1,2,3,4,5{]}. We shall
discuss about two topics combined in this paper. We would like to
briefly discuss the work that has been done in both domains before
getting into the topic. Entropy of entanglement as Quantum mechanics'
central idea is measuring the degree of entanglement {[}6{]}. In the
case of neutrino particles, the entanglement entropy represents the
degree of entanglement between the eigenstates of the neutrino mass.
Numerous theoretical investigations have tackled this matter, given
the significance of different facets of entanglement entropy in neutrino
oscillations. The literature demonstrates that Blasano et al. conducted
a number of investigations on quantum entanglement in neutrino oscillations.
As an example, Blasone et al.I've researched quantum entanglement
in neutrino oscillations in a number of ways. For example, the amount
and distribution of entanglement in the physically significant examples
of flavour mixing in quark and neutrino systems were estimated in
detail by Blasone et al {[}7{]}. The expression of mode entanglement
in terms of flavour transition probability was demonstrated in 2009
{[}8{]}. 2010 saw the investigation of single-particle entanglement
resulting from two-flavor neutrino mixing, initially in the framework
of quantum field theory and subsequently in the setting of quantum
mechanics {[}9{]}. Neutrino oscillations were explained in terms of
the (dynamical) entanglement of neutrino flavour modes in 2013 {[}10{]}.
They expanded their research into the context of quantum field theory
in 2014 and employed concurrence as a useful entanglement measure
{[}11{]}. In a subsequent study, they examined the entanglement in
flavour neutrino and antineutrino states within the context of quantum
information theory and found a correlation with experimentally measurable
quantities like lepton number and charge variances {[}12{]}. In 2015,
they studied the behaviour of multipartite entanglement within the
(three-qubit) Hilbert space of flavour neutrino eigenstates {[}13{]}.
They did this by using two measures: the concurrence and the logarithmic
negativity. In a similar vein, Kayser et al. investigated how entanglement
functions in theoretical models.Martin et al {[}14{]} investigated
measurements of quantum information, including entanglement entropy
and purity in quantum neutrino oscillations at macroscopic scales
in extreme astrophysical environments, such as the early universe,
core-collapse supernovae, and merging neutron stars. Siwach et al
examined entanglement in neutrino systems by quantifying entropies
and polarization vector components in the context of three-flavor
oscillations {[}14{]}. Amol et al presented a significant connection
between the entanglement entropy of individual neutrinos and the occurrence
of spectral splits in their energy spectra due to collective neutrino
oscillations {[}15{]}. Mallick et al explored the entanglement entropy
between spins and position space during neutrino propagation {[}16{]}.
Cervia et al measured the entanglement entropy and the Bloch vector
of the reduced density matrix, which are used to assess the interactions
between constituent neutrinos in the many-body system {[}17{]}. 

Based on Ref.{[}18{]}, we will examine the entanglement entropy between
neutrino eigenstates in matter medium. We shall demonstrate that,
in dependence on the baseline length per energy L/E and quantum entanglement
between matter dependent eigenstates. The structure of this work is
as follows: We examine the neutrino flavour oscillation model in Section
II before delving into the idea of entanglement entropy in neutrino
oscillations. Next, we apply Von Neumann's approach to compute the
entropy of neutrino oscillations between two flavours. In Section
III, we report how the modified neutrino oscillation parameter and
entanglement entropy of neutrino oscillations are affected by different
levels of baseline per energy L/E. Ultimately, a discussion and conclusion
are provided in the final part.

\section{Two Flavour Neutrino Oscillation in Vacuum and Matter}

We consider only two flavor neutrino oscillations. The two mass eigenstates,
$\nu_{1}$ and $\nu_{2}$, are linear combinations of the flavour
states $\nu_{e}$ and $\nu_{\mu}$.

\[
\nu_{e}=cos\theta\nu_{1}+sin\theta\nu_{2}
\]

\[
\nu_{\mu}=-sin\theta\nu_{1}+cos\theta\nu_{2}
\]

$\nu_{e}\rightarrow\nu_{\mu}$ neutrino oscillation probability in
long baseline neutrino experiment is influenced by the transmission
of neutrinos through the substance of earths crust. The oscillation
likelihood of matter effects {[}19, 20, 21, 22, 23, 24{]} is impressive..
In two flavour scheme, $\nu_{\mu}\rightarrow\nu_{e}$ oscillation
probability is given by

\begin{equation}
P_{e\mu}(oscillation)=sin^{2}2\theta_{12}sin^{2}\left(\frac{1.27\Delta_{21}L}{E}\right),
\end{equation}
In two flavour scheme, $\nu_{\mu}\rightarrow\nu_{e}$ survival probability
is given by

\begin{equation}
P_{e\mu}(survival=1-sin^{2}2\theta_{12}sin^{2}\left(\frac{1.27\Delta_{21}L}{E}\right),
\end{equation}

where $\Delta_{21}=m_{2}^{2}-m_{1}^{2}$ is in $eV^{2},$the baseline
length L is in Km. and the neutrino energy E in GeV. In the long baseline
experiment, the neutrinos propagates through earth crust,which has
constant density about 3 gm/cc. The oscillation probability with inclusion
of matter effects is given by

\begin{equation}
P_{e\mu}^{m}(oscillation)=sin^{2}2\theta_{12}^{m}sin^{2}\left(\frac{1.27\Delta_{21}^{m}L}{E}\right),
\end{equation}

The survival probability with inclusion of matter effects is given
by

\begin{equation}
P_{e\mu}^{m}(survival)=1-sin^{2}2\theta_{12}^{m}sin^{2}\left(\frac{1.27\Delta_{21}^{m}L}{E}\right),
\end{equation}

Comparing Eq.(1) and Eq.(2),we note that the mixing $\theta_{12}$and
$\Delta_{21}$are replaced by their matter dependent values $\theta_{12}^{m}$and
$\Delta_{21}^{m}$,these are given by

\begin{equation}
sin2\theta_{12}^{m}=\frac{\Delta_{21}sin2\theta_{12}}{\Delta_{21}^{m}},
\end{equation}

\begin{equation}
\Delta_{21}^{m}=\left(\left(\Delta_{21}cos2\theta_{12}-A\right)^{2}+\Delta_{21}^{2}sin^{2}2\theta_{12}\right)^{\frac{1}{2}}.
\end{equation}

The matter term A is given by

\begin{equation}
A(eV^{2})=2\sqrt{2}G_{F}N_{e}E_{\nu}=0.76\times10^{-4}\times\rho(gm/cc)\times E_{\nu}(MeV),
\end{equation}

where $N_{e}$ is the number of electrons per unit volume in the medium
and $\rho$ is the density of the medium. For anti-neutrinos, $A$
is replaced by $-A.$

\section{Entanglement Entropy for Neutrino Oscillation in Matter}

It is well known that a key instrument for measuring entanglement
between subsystems in bipartite quantum systems is von Neumann entropy.
S(\textgreek{r}), the Von Neumann entropy, provides an accurate way
to quantify entanglement by indicating how closely the subsystems
are correlated. The reduced density matrix is produced by tracing
out (or integrating over) the degrees of freedom of one of the subsystems
from the full density matrix of the entire system. The density matrix
is a mathematical representation that describes the quantum state
of a system, accounting for the probabilities of various possible
states.The definition of the Von Neumann entropy is as follows{[}25{]}:

\begin{equation}
S(\rho)=-Tr(\rho log\rho)
\end{equation}

where Tr\textgreek{r} = 1, where density matrices are normalised,
and \textgreek{r} is the density matrix of neutrino flavour states.

\begin{equation}
\rho=\mid\nu_{\alpha}(t)><\nu_{\alpha}(t)\mid
\end{equation}

The density matrix is a mathematical object that contains all of the
information about the state of the system. We confine the examination
to a pair of flavours, denoted as $\nu_{\alpha}$and $\nu_{\beta}$,
which correspond to the upper and lower parts of the iso-spin SU(2)
algebra of neutrino flavours {[}26{]}. Based on the assumption that
the neutrino occupation number of a particular flavour (mode) is a
reference quantum number, the following correspondences with two-qubit
states can be established:

\begin{equation}
\mid\nu_{\alpha}>=\mid1,0>=\left(\begin{array}{c}
0\\
1\\
0\\
0
\end{array}\right),
\end{equation}

\begin{equation}
\mid\nu_{\beta}>=\mid0,1>=\left(\begin{array}{c}
0\\
0\\
1\\
0
\end{array}\right)
\end{equation}

The states |0\textrangle{} and |1\textrangle{} indicate whether a
neutrino is present in mode \textgreek{a} or \textgreek{b}, respectively.
In a single-particle setting, element is thus formed among taste modalities. 

Stated otherwise, entanglement entropy is equivalent to{[}6{]}:

\begin{equation}
S(\rho)=\text{\textminus}(P_{Survival})log(P_{Survival})\text{\textminus}(P_{Oscillation})log(P_{Oscillation}))
\end{equation}

where P survival represents the possibility in Eq.(2) that a neutrino
will remain in its initial flavor state without changing flavor and
P oscillation is the likelihood in Eq.(1) that a neutrino will change
flavor while traveling across space. This relationship shows how the
entanglement entropy, as determined by the Von Neumann entropy, is
related to the likelihood that characterize the flavor survival and
oscillation of neutrinos. In matter entanglement entropy is

\begin{equation}
S^{matter}(\rho)=\text{\textminus}(P_{Survival}^{matter})log(P_{Survival}^{matter})\text{\textminus}(P_{Oscillation}^{matter})log(P_{Oscillation}^{matter}))
\end{equation}

where $P_{Oscillation}^{matter}$ indicates the chance in Eq.(4) that
a neutrino will remain in its initial flavor state without changing
flavor, and $P_{Survival}^{matter}$ indicates the likelihood in Eq.(3)
that a neutrino would change flavor while traveling across space.This
equation illustrates the relationship between the probability that
characterises the flavour survival and oscillation of neutrinos and
the entanglement entropy in matter, as given by the Von Neumann entropy
in Eq. 12.

\section{Numerical Results}

Based on Eq. (13), we examine the entanglement entropy to the neutrino
oscillations in matter. Neumann entropy was employed to analyze the
entanglement entropy between neutrino eigenstates in the matter. First,
we consider the survival chance of $P(\nu_{e}\rightarrow\nu_{\mu})$,
the transition probability of $P(\nu_{e}\rightarrow\nu_{\mu})$, and
the entanglement entropy for Eq. (11). We have set density$\rho$=3.0gm/cc,
$\theta_{12}=34^{o}$ {[}27{]}, and $\Delta_{21}=8.0\times10^{-5}eV^{2}$.

According to the previously described formula, flavour neutrino states
can be conceptualised as entangled super-positions of the mass qubits
$\brokenvert\nu_{i}>$ at any given moment, where the entanglement
is only dependent on the mixing angle and neutrino mass square difference.
The entanglement entropy of the neutrino for the two flavors of neutrino
oscillation in vacuum is displayed in Figure 1.

\begin{figure}
\includegraphics[scale=0.7]{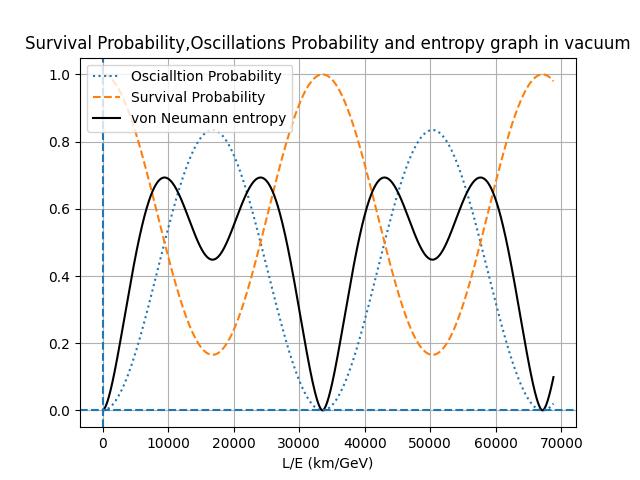}

\caption{In vaccum oscilation,survival probability and entropy}
\end{figure}

Initial state t = 0 have a clear physical interpretation: the global
state of the system is factorised, entanglement is zero, and the two
flavours are not combined. Flavours begin to fluctuate for t \ensuremath{\ge}
0, and entanglement reaches its maximum at the maximum phase.Entropy
therefore reaches its maximum at locations where there is an equal
chance of an electron neutrino surviving and a muon neutrino oscillating.
High values of entropy reflect increased mixing between the flavor
states, indicating a more entangled quantum system. Entropy once more
falls in the region between these two points, or in the range where
the probability of oscillation is larger than the probability of survival.
Entropy reaches its lowest values at the locations where the likelihood
of survival is greatest.

Now, in order to observe entropy change in the presence of anomalous
behaviour in the neutrino oscillations in matter, we quantitatively
analysed entanglement entropy using Eq. (13). We calculated entanglement
for neutrino oscillation in matter applying the Von Neumann entanglement
entropy formalism, which was previously introduced.

We now evaluated entanglement entropy using Eq. (13) in order to observe
entropy change in the presence of matter in the neutrino oscillations.
Using the previously introduced Von Neumann entanglement entropy formalism
Eq.13, we computed entanglement for constand matter density profile.
The resulting outcomes are displayed in Figs.2, 

\begin{figure}[H]
\includegraphics[scale=0.7]{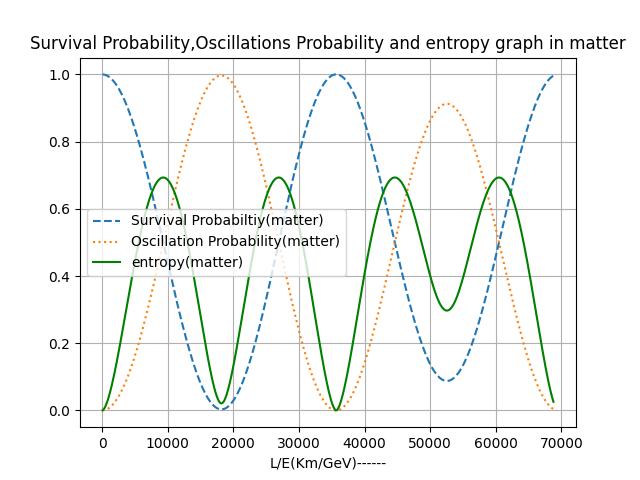}

\caption{In matter oscilation,survival probability and entropy}
\end{figure}

hi
\begin{figure}[H]
\includegraphics[scale=0.7]{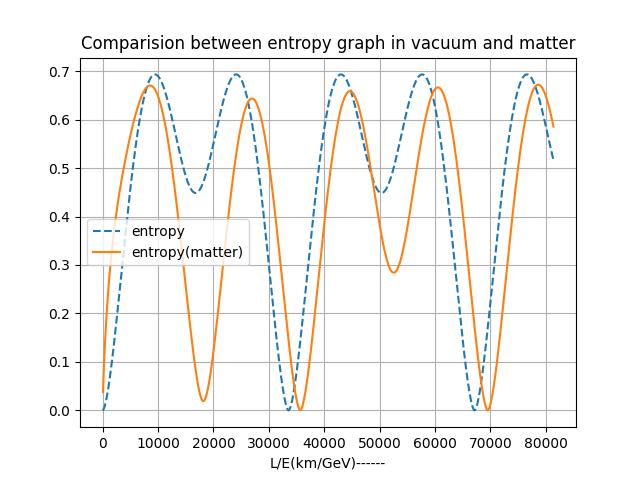}

\caption{Entropy in vaccum and matter}
\end{figure}
. The results of the entanglement entropy for the change in taste
in a matter medium were plotted. Entropy in vacuum and matter (Figure
3). It was obsreved that the matter entangle entropy area under the
L/E graph was less than the vacuum. This suggests that the degree
of entanglement is more efficient in matter. More entanglement suggests
more coherent oscillation in matter.The more effective coherent oscillation
area L/E $\leq\frac{\pi}{1.27\Delta_{21}}$ was found from the graph. 

\section{Conclusions}

In this paper,we decribe the two flavour neutrino oscillation in matter
von Neumann entropy. We relate how matter effects play a role in entanglement
and quantum correlation like von Neumann entropy. We discuss in Section
II, we provided a brief summary of matter effects on two flavour neutrino
oscillation probability. From Fig.3, we show that entanglement entropy
in matter shifts in the case of two flavour neutrino oscillation.
Numerical results indicate that due to matter effects the entanglement
entropy increases. The information entropy may increase as indicated
by the increase in the area for the entanglement entropy's subsequent
oscillation in matter. This is the outcome of the oscillation probabilities
between neutrino eigenstates experiencing a matter effects. Interpreting
the physical equivalent of the increase in entropy with each subsequent
oscillation may be premature at this point. In matter, the entropy
of a system can be describe using the von Neumann entropy, which is
define in matter as $S^{matter}(\rho)=\text{\textminus}(P_{Survival}^{matter})log(P_{Survival}^{matter})\text{\textminus}(P_{Oscillation}^{matter})log(P_{Oscillation}^{matter})).$

The amount of quantum entanglement between two subsystems in a quantum
system is expressed in terms of entanglement entropy. It was observed
that the vacuum was larger than the matter entangle entropy area under
the L/E graph. This implies that matter has a more efficient degree
of entanglement.

\end{document}